\documentclass[12pt,fleqn]{article}
\date{}
\usepackage{graphicx}
\begin{document}

\title{\bf Statistical theory of perturbation waves in transport phenomena and its experimental verification}
\author{Isaac Shnaid}
\maketitle

\begin{abstract}

In transport phenomena, perturbation waves are a result of interaction of molecules in gases and liquids, charged particles (ions, electrons) in plasma, conduction electrons and phonons in solid bodies. General statistical theory of the perturbation waves is developed and its corollaries are studied. On this basis is proved universality of introduced earlier local time concept, which leads to a formulation of kinetic, conservation and governing equations for macroscopic transport phenomena with finite speed of the perturbations propagation in gases, liquids, solids and plasma.

Speed of thermal perturbations propagation in phonon and Fermi electron  gases and plasma, and also speed of thermal, momentum and mass perturbations propagation in ideal gas are theoretically determined.

It is shown that published experimental results for femtosecond laser heating of thin gold films and results of power modulation experiments in JET tokamak agree with the developed theory. 

\end{abstract}

Keywords: \emph{Transport phenomena, perturbation waves, kinetic theory.}

\section{Introduction}

It is well known that if a heat conducting system is initially in thermal equilibrium, and a local thermal perturbation is introduced, then according to Fourier equation temperature everywhere in the system instantaneously becomes perturbed. It means that according to Fourier equation, speed of thermal perturbations propagation is infinite.  All other classical transport equations for viscous flow and diffusion also predict infinite speed of perturbations propagation. In reality, perturbation waves have finite speed, because perturbations propagation is a result of interaction of molecules in gases and liquids, charged particles (ions, electrons) in plasma, conduction electrons and phonons in solid bodies. 

In my previous works \cite{sh7a}-\cite{sh6}, approximate statistical theory of the perturbations waves in transport processes and phenomenological thermodynamic theory of thermal perturbations propagation were developed. They led to a proof of the local time concept for systems with zero or small macroscopic velocity, which leads to a formulation of kinetic, conservation and governing equations for real transport processes with finite speed of the perturbations propagation. Correctness of the local time concept was confirmed by analysis of experimental data for electron temperature pulses propagation in magnetically confined hot plasma.

In the present work, general statistical theory of the perturbation waves is developed and its corollaries are studied. From the theory follows that the local time concept is universal and is applicable  to all transport processes in non-equilibrium systems in gases, liquids, solids and plasma.

I theoretically determine speed of thermal perturbations propagation in phonon and Fermi electron gases, and speed of thermal, momentum and mass perturbations propagation in ideal gas. For plasma in strong external magnetic fields, small values of the perturbations propagation speed in direction perpendicular to the field are theoretically predicted. I also study influence of macroscopic velocity on speed of the perturbations propagation.

I show that published experimental results for femtosecond laser heating of thin gold films are in good agreement with the theory of the perturbation waves and the local time concept. I also find that experimental and calculated theoretical values of speed of thermal perturbations propagation in electron Fermi gas are close.

I prove that published results of power modulation experiments with magnetically confined hot plasma in JET tokamak confirm correctness of the theory of the perturbation waves, the local time concept, and theoretical prediction of small values of the perturbations propagation speed in direction perpendicular to the field. These experiments also support theoretical prediction of equal values of thermal perturbations propagation speed in electron and ion plasma components.

\section{Statistical formulation of the problem}

I analyse a general case of non-equilibrium systems and use term {\em particle} for molecules in neutral gases and liquids, free ions and electrons in plasma, phonons and conduction electrons in solids.  It is assumed that a {\em non-steady, non-equilibrium  state field} of thermodynamic and hydrodynamic parameters, for instance temperature, pressure, concentration and macroscopic velocity, exists initially in a region. Then some part of the initial field is perturbed. As a result, interacting perturbed and non-perturbed subregions appear in the region. Further analysis applies the same statistical method that was developed in my previous works \cite{sh7a}, \cite{sh7} where I analysed a particular case when initial field is in {\em equilibrium or steady non-equilibrium state.}

Let $E$ be an arbitrary point of the surface $S_P$ separating the perturbed subregion from the non-perturbed one.  At that point a local cartesian coordinate system $X,~Y,~Z$ is introduced, where axis $X$ is normal to the surface $S_P$ at the point $E$, and axes $Y$ and $Z$ create a plane which is tangential to the surface $S_P$. The coordinate system $X,~Y,~Z$ does not move with respect to the non-perturbed subregion. The axis $X$ has direction from the perturbed subregion to the non-perturbed one. For the perturbed and the non-perturbed subregions distribution functions for selected kind of particles are designated as  $f({\bf r},{\bf v},\tau)$ and $f_0({\bf r},{\bf v}, \tau)$, respectively, where ${\bf r}(X,Y,Z)$ and ${\bf v}(v_X,v_Y,v_Z)$ denote the particle position vector and  particle velocity vector, respectively; $\tau$ is time. At the point $E$, particles with positive $v_X$ velocity component $v_X=v_X^{(+)}>0$ enter the non-perturbed subregion from the perturbed subregion, and particles with negative $v_X$ velocity component $v_X=v_X^{(-)}<0$ enter the perturbed subregion from the non-perturbed one. If $F$ is distribution function at the point $E$,  there is $F=f({\bf r},{\bf v},\tau)$ for particles with $v_X=v_X^{(+)}>0$, and $F=f_0({\bf r},{\bf v},\tau)$ for particles with $v_X=v_X^{(-)}<0$. Here and further superscript $(+)$ is used for particles with positive $v_X=v_X^{(+)}$ velocity component, and superscript $(-)$ - for particles with negative $v_X=v_X^{(-)}$ velocity component.

Let $\varphi_S$ be a property of a particle crossing the surface $S_P$, $\varphi({\bf r},{\bf v},\tau)$ is the same property of a particle in the perturbed subregion, while $\varphi_0({\bf r},{\bf v},\tau)$ is the same property in the non-perturbed subregion. Thus, $\varphi_S=\varphi({\bf r},{\bf v},\tau)$ for particles with $v_X=v_X^{(+)}>0$, and $\varphi_S=\varphi_0({\bf v})$ when $v_X=v_X^{(-)}<0$.

The distribution function $F$ satisfies general kinetic Boltzmann equation applied for selected kind of particles
\begin{equation}
\frac{\partial F}{\partial \tau} + v_X \frac {\partial F}{\partial X}+v_Y \frac {\partial F}{\partial Y}+v_Z \frac {\partial F}{\partial Z}=D_C-G_M
\label{eq:1101}
\end{equation}
where $D_C$ is a collision term; $G_M$ is  a momentum term
\begin{equation}
G_M=\frac {\partial p_X}{\partial \tau} \frac {\partial F}{\partial p_X}+\frac {\partial p_Y}{\partial \tau} \frac {\partial F}{\partial p_Y}+\frac {\partial p_Z}{\partial \tau} \frac {\partial F}{\partial p_Z}
\label{eq:1101ab}
\end{equation}
and ${\bf p}(p_X,p_Y,p_Z)$ is a particle momentum. We do not introduce any limiting assumption regarding structure of the collision and momentum terms in equation (\ref{eq:1101}).

As usually \cite{sh4}, both sides of equation (\ref{eq:1101}) are multiplied by $\varphi_S$ and after simple transformations it becomes

\begin{displaymath}
\frac{\partial (\varphi_S F) }{\partial \tau} + v_X \frac {\partial (\varphi_S F)}{\partial X}+v_Y \frac { \partial (\varphi_S F)}{\partial Y}+v_Z \frac { \partial (\varphi_S f)}{\partial Z}=
\end{displaymath}
\begin{equation}
=\varphi_S (D_C-G_M) + F~\bigg ( \frac{\partial \varphi_S}{\partial \tau} + v_X \frac {\partial \varphi_S}{\partial X}+v_Y \frac {\partial \varphi_S}{\partial Y}+v_Z \frac {\partial \varphi_S}{\partial Z} \bigg )
\label{eq:1101d}
\end{equation}

All terms of expression (\ref{eq:1101d}) are  integrated over velocities ${\bf v}$, and integrals  are transformed taking into account formulated earlier properties of functions $F$ and $\varphi_S$. It is suggested that for all values of velocity components $v_X^{(+)},~v_Y,~v_Z$ derivatives $\frac{\partial (\varphi f)}{\partial X},~\frac{\partial (\varphi f)}{\partial Y},~\frac{\partial (\varphi f)}{\partial Z}$ do not change the signs, and for all values of velocity components $v_X^{(-)},~v_Y,~v_Z$ derivatives $\frac{\partial (\varphi_0 f_0)}{\partial X},~\frac{\partial (\varphi_0 f_0)}{\partial Y},~\frac{\partial (\varphi_0 f_0)}{\partial Z}$ also do not change the signs. Therefore the integral mean value theorem \cite{sh37} is applicable. After integration the following expressions are obtained:

\begin{equation}
I_{\tau}+I_X+I_Y+I_Z=I_{CM} + I_S
\label{eq:1102}
\end{equation}

\begin{displaymath}
I_{\tau}=\int  \frac{\partial (\varphi_S F)}{\partial \tau}~dv_X~dv_Y~dv_Z=
\end{displaymath}
\begin{displaymath}
=\int \frac{\partial (\varphi f)}{\partial \tau}~dv_X^{(+)}~dv_Y~dv_Z +\int  \frac{\partial (\varphi_0 f_0)}{\partial \tau}~dv_X^{(-)}~dv_Y~dv_Z=
\end{displaymath}
\begin{equation}
=\frac {\partial (n^{(+)}\overline{\varphi}^{(+)})}{\partial \tau}+\frac {\partial (n^{(-)}\overline{\varphi}_0^{(-)})}{\partial \tau}
\label{eq:1103a}
\end{equation}

\begin{displaymath}
I_X=\int v_X \frac{\partial (\varphi_S F)}{\partial X}~dv_X~dv_Y~dv_Z=
\end{displaymath}
\begin{displaymath}
=\overline{v_X^{(+)}} \int \frac{\partial (\varphi f)}{\partial X}~dv_X^{(+)}~dv_Y~dv_Z+ \overline{v_X^{(-)}} \int \frac{\partial (\varphi_0 f_0)}{\partial X}~dv_X^{(-)}~dv_Y~dv_Z=
\end{displaymath}
\begin{equation}
=\overline{v_X^{(+)}}~\frac {\partial (n^{(+)}\overline{\varphi}^{(+)})}{\partial X}+\overline{v_X^{(-)}} \frac {\partial (n^{(-)}\overline{\varphi_0}^{(-)})}{\partial X}
\label{eq:1104a}
\end{equation}

\begin{displaymath}
I_Y=\int v_Y \frac{\partial (\varphi_S F)}{\partial Y}~dv_X~dv_Y~dv_Z=
\end{displaymath}
\begin{displaymath}
=\overline{v_Y}^{(+)}  \int \frac{\partial (\varphi f)}{\partial Y}~dv_X^{(+)}~dv_Y~dv_Z+\overline{v_Y}^{(-)} \int \frac{\partial (\varphi_0 f_0)}{\partial Y}~dv_X^{(-)}~dv_Y~dv_Z =
\end{displaymath}
\begin{equation}
=\overline{v_Y}^{(+)}~\frac {\partial (n^{(+)}\overline{\varphi}^{(+)})}{\partial Y}+\overline{v_Y}^{(-)} \frac {\partial (n^{(-)}\overline{\varphi_0}^{(-)})}{\partial Y}
\label{eq:1105a}
\end{equation}

\begin{displaymath}
I_Z=\int v_Z  \frac{\partial (\varphi_S F)}{\partial Z}~dv_X~dv_Y~dv_Z=
\end{displaymath}
\begin{displaymath}
=\overline{v_Z}^{(+)} \int \frac{\partial (\varphi f)}{\partial Z}~dv_X^{(+)}~dv_Y~dv_Z+\overline{v_Z}^{(-)} \int \frac{\partial (\varphi_0 f_0)}{\partial Z}~dv_X^{(-)}~dv_Y~dv_Z =
\end{displaymath}
\begin{equation}
=\overline{v_Z}^{(+)}~\frac {\partial (n^{(+)}\overline{\varphi}^{(+)})}{\partial Z}+\overline{v_Z}^{(-)} \frac {\partial (n^{(-)}\overline{\varphi_0}^{(-)})}{\partial Z}
\label{eq:1106a}
\end{equation}

\begin{equation}
I_{CM}=\int \varphi_S (D_C-G_M)~dv_X~dv_Y~dv_Z
\label{eq:1107}
\end{equation}

\begin{displaymath}
I_S=\int F~\bigg ( \frac{\partial \varphi_S}{\partial \tau} + v_X \frac {\partial \varphi_S}{\partial X}+v_Y \frac {\partial \varphi_S}{\partial Y}+v_Z \frac {\partial \varphi_S}{\partial Z} \bigg )~dv_X~dv_Y~dv_Z=
\end{displaymath}
\begin{displaymath}
=\int f~\bigg ( \frac{\partial \varphi}{\partial \tau} + v_X^{(+)} \frac {\partial \varphi}{\partial X}+v_Y \frac {\partial \varphi}{\partial Y}+v_Z \frac {\partial \varphi}{\partial Z} \bigg )~dv_X^{(+)}~dv_Y~dv_Z +
\end{displaymath}
\begin{displaymath}
+\int f_0~\bigg ( \frac{\partial \varphi_0}{\partial \tau} + v_X^{(-)} \frac {\partial \varphi_0}{\partial X}+v_Y \frac {\partial \varphi_0}{\partial Y}+v_Z \frac {\partial \varphi_0}{\partial Z} \bigg )~dv_X^{(-)}~dv_Y~dv_Z=
\end{displaymath}
\begin{equation}
=n^{(+)}~\overline {\bigg ( \frac{\partial \varphi}{\partial \tau} + v_X \frac {\partial \varphi}{\partial X}+v_Y \frac {\partial \varphi}{\partial Y}+v_Z \frac {\partial \varphi}{\partial Z} \bigg )}^{(+)}+n^{(-)}~\overline {\bigg ( v_X \frac {\partial \varphi_0}{\partial X}+v_Y \frac {\partial \varphi_0}{\partial Y}+v_Z \frac {\partial \varphi_0}{\partial Z} \bigg )}^{(-)}
\label{eq:1108h}
\end{equation}
where $n$ is mean number of particles per unit volume, overline denotes mean value of an appropriate parameter, and

\begin{equation}
\overline{v_X^{(+)}}=\int v_X^{(+)}\frac{\partial (\varphi f)}{\partial X}~dv_X^{(+)}~dv_Y~dv_Z \bigg / \int \frac{\partial (\varphi f)}{\partial X}~dv_X^{(+)}~dv_Y~dv_Z=c_X
\label{eq:1108i}
\end{equation}

\begin{equation}
\overline{v_X^{(-)}}=\int v_X^{(-)}\frac{\partial (\varphi_0 f_0)}{\partial X}~dv_X^{(-)}~dv_Y~dv_Z \bigg / \int \frac{\partial (\varphi_0 f_0)}{\partial X}~dv_X^{(-)}~dv_Y~dv_Z=c_{0X}
\label{eq:1108ia}
\end{equation}

\begin{equation}
\overline{v_Y^{(+)}}=\int v_Y\frac{\partial (\varphi f)}{\partial Y}~dv_X^{(+)}~dv_Y~dv_Z \bigg / \int \frac{\partial (\varphi f)}{\partial Y}~dv_X^{(+)}~dv_Y~dv_Z=c_Y
\label{eq:1108k}
\end{equation}

\begin{equation}
\overline{v_Y^{(-)}}=\int v_Y\frac{\partial (\varphi_0 f_0)}{\partial Y}~dv_X^{(-)}~dv_Y~dv_Z \bigg / \int \frac{\partial (\varphi_0 f_0)}{\partial Y}~dv_X^{(-)}~dv_Y~dv_Z=c_{0Y}
\label{eq:1108ka}
\end{equation}

\begin{equation}
\overline{v_Z^{(+)}}=\int v_Z\frac{\partial (\varphi f)}{\partial Z}~dv_X^{(+)}~dv_Y~dv_Z \bigg / \int \frac{\partial (\varphi f)}{\partial Z}~dv_X^{(+)}~dv_Y~dv_Z=c_Z
\label{eq:1108l}
\end{equation}

\begin{equation}
\overline{v_Z^{(-)}}=\int v_Z\frac{\partial (\varphi_0 f_0)}{\partial Z}~dv_X^{(-)}~dv_Y~dv_Z \bigg / \int \frac{\partial (\varphi_0 f_0)}{\partial Z}~dv_X^{(-)}~dv_Y~dv_Z=c_{0Z}
\label{eq:1108la}
\end{equation}
denote respective mean integral values of $v_X^{(+)},~v_X^{(-)},~v_Y^{(+)},~v_Y^{(-)},~v_Z^{(+)},~v_Z^{(-)}$. By definition, $c_X>0,~c_{0X}<0$, and $c_Y,~c_{0Y},~c_Z,~c_{0Z}$ may have any signs or to be zero.

\section{Evolution of the perturbed and non-perturbed subregions}

After substitution  of formulas (\ref{eq:1103a})-(\ref{eq:1108h}) in (\ref{eq:1102}) we obtain expressions describing evolution of the perturbed and non-perturbed subregions

\begin{equation}
\frac{D \phi}{D \tau}+\frac{D \phi_0}{D \tau} =A(\tau,X,Y,Z)
\label{eq:1109g}
\end{equation}
\begin{equation}
\frac{D \phi}{D \tau}=\frac {\partial \phi}{\partial \tau} +c_X~\frac {\partial \phi}{\partial X}+c_Y~\frac {\partial \phi}{\partial Y}+c_Z~\frac {\partial \phi}{\partial Z}
\label{eq:1109ga}
\end{equation}
\begin{equation}
\frac{D \phi_0}{D \tau}=\frac {\partial \phi_0}{\partial \tau} +c_{0X}~\frac {\partial \phi_0}{\partial X}+c_{0Y}~\frac {\partial \phi_0}{\partial Y}+c_{0Z}~\frac {\partial \phi_0}{\partial Z}
\label{eq:1109gb}
\end{equation}
\begin{equation}
A(\tau,X,Y,Z)=I_{CM}+ I_S
\label{eq:1109gc}
\end{equation}
where $\phi(\tau, X, Y, Z)=n^{(+)}\overline{\varphi}^{(+)},~c_X(\tau, X, Y, Z)>0,~c_Y(\tau, X, Y, Z),~c_Z(\tau, X, Y, Z)$ are parameters determined only by averaged properties of the perturbed subregion, while  $\phi_0(\tau, X, Y, Z)=n^{(-)}\overline{\varphi}_0^{(-)},~c_{0X}(\tau, X, Y, Z)<0,~c_{0Y}(\tau, X, Y, Z),~c_{0Z}(\tau, X, Y, Z)$ are defined only by averaged properties of the non-perturbed subregion.

Macroscopic formulae (\ref{eq:1109g})-(\ref{eq:1109gc}) directly follow from the general kinetic Boltzmann equation (\ref{eq:1101}). They include total derivatives $\frac{D \phi}{D \tau}$ and $\frac{D \phi_0}{D \tau}$ describing evolution of the boundary surface $S_P$ separating the perturbed and non-perturbed subregions. As the perturbed subregion is in non-steady non-equilibrium state, therefore in formula (\ref{eq:1109ga})  both local $\frac {\partial \phi}{\partial \tau} \neq 0$ and convective  $c_X~\frac {\partial \phi}{\partial X}+c_Y~\frac {\partial \phi}{\partial Y}+c_Z~\frac {\partial \phi}{\partial Z} \neq 0$ derivatives are present. It means that total derivative $\frac{D \phi}{D \tau}\neq 0$ describes propagation of the perturbed subregion into the non-perturbed one with finite total speed ${\bf c}(c_X, c_Y, c_Z)$ and normal speed $c_X>0$. To prove it we multiply both sides of expression (\ref{eq:1109ga}) by $\delta \tau$ which is a small time interval, and obtain following formula
\begin{displaymath}
\Delta \phi= \phi(\tau+\delta \tau, X+c_X~\delta \tau, Y+c_Y~\delta \tau, Z+c_Z~\delta \tau)-\phi(\tau, X, Y, Z)= \frac{D \phi}{D \tau}~\delta \tau=
\end{displaymath}
\begin{equation}
=\bigg(\frac {\partial \phi}{\partial \tau} +c_X~\frac {\partial \phi}{\partial X}+c_Y~\frac {\partial \phi}{\partial Y}+c_Z~\frac {\partial \phi}{\partial Z}\bigg)~\delta \tau
\label{eq:1109i}
\end{equation}

Formula (\ref{eq:1109i}) shows that if $X, Y, Z$ are coordinates of a point, belonging to the perturbed subregion  and located at the initial time moment $\tau$  on the surface $S_P$, which separates the perturbed subregion from the non-perturbed one, at the next time moment $\tau+\delta \tau$  the field in a point with coordinates $X+c_X~\delta \tau,~ Y+c_Y~\delta \tau,~ Z+c_Z~\delta \tau$ becomes perturbed, while at initial time moment $\tau$ it was not perturbed. It means that the boundary of the perturbed subregion gradually propagates in the non-perturbed subregion with finite normal mean speed $c_X(\tau, X, Y, Z)$. Therefore $c_X(\tau, X, Y, Z)$ is local mean speed of the perturbations propagation.

In the same way we can prove that if the non-perturbed subregion is in non-steady non-equilibrium state,  the total derivative $\frac{D \phi_0}{D \tau} \neq 0$ describes propagation of the non-perturbed subregion into the perturbed one with finite total speed ${\bf c_0}(c_{0X}, c_{0Y}, c_{0Z})$ and normal speed $c_{0X}<0$. {\em Therefore in this case, at initial moment of time, when the perturbation was introduced, the boundary surface $S_P$ instantaneously disappears, two subregions - perturbed and non-perturbed - join, and the perturbations propagation is principally non-observable.}

If the non-perturbed subregion is in steady non-equilibrium state, local derivative $\frac {\partial \phi_0}{\partial \tau}=0$ disappears in formula (\ref{eq:1109gb}), and only convective derivative $c_{0X}~\frac {\partial \phi_0}{\partial X}+c_{0Y}~\frac {\partial \phi_0}{\partial Y}+c_{0Z}~\frac {\partial \phi_0}{\partial Z}\neq 0$ is left. Now the total derivative $\frac{D \phi_0}{D \tau}$ does not describe propagation of the non-perturbed steady state subregion into the perturbed one, because such propagation must be accompanied by change of respective properties in time, and propagation of steady state field is physically impossible. In a particular case of equilibrium in the non-perturbed subregion there is $\frac{D \phi_0}{D \tau}=0$.
 
\emph {Therefore we proved that in all non-equilibrium processes the perturbations propagation is observable and physically meaningful only when the perturbed subregion interacts with steady state non-perturbed subregion, and only propagation of the non-steady state perturbed subregion into the steady state non-equilibrium or equilibrium non-perturbed subregion is physically possible. It means, that in all cases of gas, liquid, solid  and plasma non-equilibrium systems, the non-equilibrium perturbed subregion of any three-dimensional shape gradually replaces the non-perturbed steady state subregion with local normal speed $c_X=\overline{v_X}^{(+)}>0$ which is speed of the perturbations propagation defined by expression (\ref{eq:1108i}).}

By definition, speed of the perturbations propagation $c_X>0$ is positive  and finite.  The limitations introduced by the integral mean value theorem \cite{sh37} ensure that for any arbitrary distribution function $f$ and particle property $\varphi$, expression (\ref{eq:1108i}) determines positive, finite and physically sound values of speed of the perturbations propagation. If these limitations do not hold, the expression (\ref{eq:1108i}) does not define positive and finite values of $c_X$, and does not serve more as statistical definition of  the perturbations propagation speed. In such case, other methods of $c_X$ determination can be applied, for instance, numerical statistical modelling of the perturbed subregion propagation or experimental study of the perturbations propagation. When the integral mean value theorem does not hold, approximate values of speed  of the perturbations propagation can be calculated if instead of derivative $\frac{\partial (\varphi f)}{\partial X}$ to use its absolute value $|\frac{\partial (\varphi f)}{\partial X}|$
\begin{equation}
c_X=\overline{v_X^{(+)}}=\int v_X^{(+)} \bigg |\frac{\partial (\varphi f)}{\partial X}\bigg |~dv_X^{(+)}~dv_Y~dv_Z \bigg / \int \bigg |\frac{\partial (\varphi f)}{\partial X}\bigg |~dv_X^{(+)}~dv_Y~dv_Z
\label{eq:1110}
\end{equation}

When $\frac{\partial (\varphi f)}{\partial X}$ does not change its sign during integration, formula (\ref{eq:1110}) is identical to (\ref{eq:1108i}) and defines theoretically exact value of $c_X$, in the opposite case it defines approximate value of speed of the perturbations propagation, whose accuracy is checked later by comparison with experimental data for thermal perturbations propagation in Fermi electron gas. Thus, expression (\ref{eq:1110}) is universal, and we will always use it for theoretical determination of speed of the perturbations propagation.

\section{Perturbation traveltime and generalization of the local time concept}

As in my previous works \cite{sh7a}-\cite{sh6}, I assume that perturbation traveltime $\tau_P(x,y,z)$ defines a time moment when the perturbation reached a given point $M(x,y,z)$,  where $x,~y,~z$ is an arbitrary cartesian coordinate system not moving with respect to the non-perturbed steady state subregion. Therefore the boundary surface $S_P$, separating the perturbed from the non-perturbed steady state subregion, is a surface of constant perturbation traveltime  $\tau_P(x,y,z)=const$. From definition of perturbation traveltime follows

\begin{equation}
|\nabla \tau_P|=\frac {1}{c_X(\tau_P, x, y, z)}
\label{eq:1111b}
\end{equation}
or
\begin{equation}
\bigg ( \frac {\partial \tau_P}{\partial x}\bigg )^2+\bigg ( \frac {\partial \tau_P}{\partial y}\bigg )^2+\bigg ( \frac {\partial \tau_P}{\partial z}\bigg )^2=\frac{1}{c_X^2(\tau_P, x, y, z)}
\label{eq:1111}
\end{equation}

In the equation (\ref{eq:1111}) instead of speed $c_X(\tau_P, x, y, z)$ can be used slowness
\begin{equation}
s_X(\tau_P, x, y, z)=\frac{1}{c_X(\tau_P, x, y, z)}
\label{eq:1112}
\end{equation}

The non-linear equation (\ref{eq:1111}) is similar to the classical eikonal equation \cite{sh38}, \cite{sh39} and determines traveltime in the case of any macroscopic velocity. According to results obtained in previous sections, only its solutions corresponding to primary perturbation waves propagating in the non-perturbed steady state subregion, have physical meaning.

From equation ({\ref{eq:1111}) follows that
\begin{equation}
c_X(\tau_P, x, y, z)=c_X(\tau_P(x,y,z), x, y, z)=c_X(x, y, z)
\label{eq:1112a}
\end{equation}
\begin{equation}
s_X(\tau_P, x, y, z)=s_X(\tau_P(x,y,z), x, y, z)=s_X(x, y, z)
\label{eq:1112b}
\end{equation}
are functions only of coordinates.

In a general case, the non-perturbed steady state subregion can be perturbed in different places and at different time moments. Accordingly solution of eikonal equation ({\ref{eq:1111}) must satisfy several initial conditions of the type
\begin{equation}
\tau_{0P}(x,y,z)=\tau_0
\label{eq:1112ba}
\end{equation}
where $\tau_{0P}(x,y,z)$ and $\tau_0$ denote initial value of traveltime and time moment when the perturbation was introduced, respectively.  

\emph{ So if the non-perturbed initial steady state subregion was perturbed in different places and at different time moments, perturbed non-steady state subregion is permanently divided on different sub-subregions corresponding to different initial conditions of type (\ref{eq:1112ba})}.

For any point $M(x,y,z)$ of the region, local time is determined as time counted from a moment when the perturbation reached the point \cite{sh7a}-\cite{sh6}
\begin{equation}
\vartheta(x,y,z,\tau)=\tau-\tau_P(x,y,z)
\label{eq:1114}
\end{equation}
For this point at a global time moment $\tau$, three characteristic cases exist:

\begin{enumerate}
\item $\vartheta=\tau-\tau_P<0$, the perturbation has not reached the point. Thus, the point belongs to the non-perturbed steady state subregion.
\item $\vartheta=\tau-\tau_P=0$, the perturbation has reached the point $M$. Now $M$ is located on the border surface $S_P$, separating the perturbed and the non-perturbed subregions.
\item $\vartheta=\tau-\tau_P>0$, the point $M$ is inside the perturbed subregion, where transport processes occur. All further considerations are formulated for the perturbed subregion where $\vartheta \geq 0$.
\end{enumerate}

For a case when the perturbed subregion interacts with steady state non-perturbed subregion, fundamental equation (\ref{eq:1109g}) has a form
\begin{equation}
\frac {\partial \phi}{\partial \tau} +c_X~\frac {\partial \phi}{\partial X}+c_Y~\frac {\partial \phi}{\partial Y}+c_Z~\frac {\partial \phi}{\partial Z}=
=A_0(\tau,X,Y,Z)
\label{eq:1113a}
\end{equation}
where $A_0=A-(c_{0X}~\frac {\partial \phi_0}{\partial X}+c_{0Y}~\frac {\partial \phi_0}{\partial Y}+c_{0Z}~\frac {\partial \phi_0}{\partial Z})$. In a case of non-zero macroscopic velocity there is $c_Y \neq 0,~c_Z \neq 0$ in expression (\ref{eq:1113a}). 

Further in formula (\ref{eq:1113a}) we can replace independent variables $\tau,~X,~Y,~Z$ by $\vartheta,~X,~Y,~Z$. As
\begin{displaymath}
\frac{\partial \phi}{\partial \tau}=\frac{\partial \phi}{\partial \vartheta},~~~
\frac{\partial \phi}{\partial X}\bigg | _\tau=\frac{\partial \phi}{\partial X}\bigg | _\vartheta - \frac {1}{c_X}~\frac{\partial \phi}{\partial \vartheta},~~~
\frac{\partial \phi}{\partial Y}\bigg | _\tau=\frac{\partial \phi}{\partial Y}\bigg | _\vartheta,~~~
\frac{\partial \phi}{\partial Z}\bigg | _\tau=\frac{\partial \phi}{\partial Z}\bigg | _\vartheta,
\end{displaymath}
equation (\ref{eq:1113a}) becomes
\begin{equation}
c_X \frac{\partial \phi}{\partial X}\bigg | _\vartheta+c_Y \frac{\partial \phi}{\partial Y}\bigg | _\vartheta+c_Z \frac{\partial \phi}{\partial Z}\bigg | _\vartheta={\bf c} \cdot \nabla \phi= A_0
\label{eq:1113aa}
\end{equation}
where symbols $|_\tau,~|_\vartheta$ mean that independent variables are $\tau,~X,~Y,~Z$ or $\vartheta,~X,~Y,~Z$, respectively.

Finally we get
\begin{equation}
\frac{\partial \phi}{\partial X_c}\bigg | _\vartheta= \frac {1}{c}~A_0
\label{eq:1113ab}
\end{equation}
where coordinate $X_c$ is measured along axis parallel to vector ${\bf c}(c_X,~c_Y,~c_Z)$, and $c=\sqrt{c_X^2+c_Y^2+c_Z^2}$.

Left side of the general fundamental equation (\ref{eq:1113a}) includes first order time and space derivatives and explicitly depends on velocity components $c_X, c_Y, c_Z$. Expression (\ref{eq:1113ab}) is fully equivalent to equation (\ref{eq:1113a}), but its left side does not depend explicitly on $c_X, c_Y, c_Z$, and speed of the perturbations propagation $c_X$  affects only local time $\vartheta$. Because of this, transformed equation (\ref{eq:1113ab}) is identical for finite $c_X$, when $\tau_P > 0$, and for the classical case of infinite $c_X$, when always $\tau_P = 0$. The only difference is: for finite $c_X$ there is $0<\vartheta < \tau$, while for the classical case always $\vartheta = \tau$. Therefore when independent variables $\vartheta,~X,~Y,~Z$ are used in all transport equations, the same equations describe the classical case of infinite speed of the perturbations propagation when $\vartheta = \tau$, and the case of finite speed of the perturbations propagation when $0<\vartheta < \tau$. Thus, we proved that the local time concept, formulated in my previous works \cite{sh7a}-\cite{sh6} for systems with zero macroscopic velocities $c_Y=0,~c_Z=0$, is universal and holds for a general case of non-zero macroscopic velocities when $c_Y \neq 0,~c_Z \neq 0$. It means that:

\emph{For all non-equilibrium systems in gases, liquids, solids and plasma, any macroscopic velocities and all transport processes, kinetic, conservation and governing equations in a case of finite speed  of the perturbations propagation and the respective classical  equations with infinite speed  of the perturbations propagation (for instance, Fourier, Fick, Euler, Navier-Stokes and magnetohydrodynamic equations) \underline {are identical}, if independent variables  $\vartheta=\tau-\tau_P,~x,~y,~z$ are used. The same equations become classical when $\tau_P=0,~ \vartheta=\tau$ that corresponds to $c_X \rightarrow \infty$, and they describe the finite $c_X$ case  when $\tau_P > 0,~ 0 < \vartheta < \tau$. Because of this, all equations written in such a form are called \underline {modified  equations}.}.

For classical transport processes such as heat conduction, diffusion and viscous flow, modified equations and some analytical solutions of modified governing equations are presented in papers \cite{sh7a}-\cite{sh6}.

Immediate corollaries of the local time concept are \cite{sh7a}-\cite{sh6}:

\begin{enumerate}
\item Modified governing equations for all transport processes with finite speed of the perturbations propagation are of the same parabolic type, as classical equations, thus, fully excluding any possibility of reversing the transport process. These equations are consistent with the Second Law of thermodynamics because their solutions satisfy the Maximum Principle. According to the Maximum Principle for parabolic partial differential equations without a source term, maximum value of any scalar characteristic parameter, e.g. temperature, is reached  either on boundary of the domain or at the initial moment of time.

Numerous earlier attempts of many authors to introduce transport equations with finite speed of the perturbations propagation were not successful because they tried to formulate hyperbolic governing transport equations which are not consistent with thermodynamics, and therefore are non-physical. For instance, some their solutions describe flow of heat from cold to hot bodies in contradiction with the Second Law of thermodynamics. Full enough list of these works and their analysis I give in papers \cite{sh7a}-\cite{sh6}.

\item Let  $\beta(x,y,z,\tau)$ be a solution of a classical transport governing equation, satisfying certain boundary conditions and an initial condition $\beta(x,y,z,0)=\beta_0 (x, y, z)$, where $\beta$ is a scalar characteristic parameter.  According to the local time concept, a solution of the modified governing transport equation is the same function but with local time as an argument $\beta(x,y,z,\vartheta)$.

    Solutions of classical parabolic type governing transport equations predict that any local perturbation introduced at initial  moment of time $\tau_i=0$, instantaneously affects all space domain. The modified parabolic governing equations use local time $\vartheta$ as an independent variable. In this case, initial perturbation introduced at initial local time moment $\vartheta_i=0$,  affects the space domain at the same local time value $\vartheta_i=\tau_{im}-\tau_P(x,y,z)=0$.  Therefore perturbation arrives to an arbitrary point $M(x,y,z)$ at global time moment $\tau_{im}=\tau_P(x,y,z)$, i.e. with global time delay. So the modified parabolic governing equations predict finite speed of the perturbations propagation, and the perturbations have wave behaviour. As a result of transport processes irreversibility, these waves cannot reflect and interfere because only solutions of eikonal  equation (\ref{eq:1111}) corresponding to primary perturbation waves propagating in the non-perturbed subregion, have physical meaning.

On the  boundary of the perturbed subregion it is $\vartheta = 0$, and therefore on the boundary is $\beta(x,y,z,\vartheta)=\beta_0 (x, y, z)$. Therefore the propagating wave of any characteristic parameter does not have a sharp front.

\item From formulae derived in previous paragraph follows that solutions of classical transport governing equations are accurate enough when the transport process is slow and/or perturbation traveltime is small enough.  To prove this, we choose an arbitrary point with coordinates $x_A,~y_A,~z_A$ and perturbation traveltime value $\tau_{PA}(x_A,y_A,z_A)$, where there is $\beta(x_A,y_A,z_A,\vartheta=\tau-\tau_{PA})=\beta_0 (x, y, z)$ at initial moment of time $\tau \leq \tau_{PA}$. Then becomes $\tau > \tau_{PA}$, and $\beta(x_A,y_A,z_A,\vartheta=\tau-\tau_{PA})\neq \beta_0 (x, y, z)$. Obviously, the difference between solutions of classical and modified governing transport equations for the perturbed subregion 
\begin{equation}
|\Delta \beta|=|\beta(x_A,y_A,z_A,\tau)-\beta(x_A,y_A,z_A,\vartheta=\tau-\tau_{PA})| \approx \bigg |\frac {\partial \beta}{\partial \tau} \bigg |~ \tau_{PA}
\label{eq:1117a}
\end{equation}
is small for small values of perturbation traveltime $\tau_{PA}$ and/or slow processes, when $|\frac {\partial \beta}{\partial \tau}|$ is small.

\end{enumerate}

\section{Modified heat conduction equations}

As an illustration, the modified Fourier kinetic, conservation and governing equations for heat conduction with finite speed of heat propagation in isotropic continuum are reproduced from \cite{sh7a}-\cite{sh6}

\begin{equation}
{\bf q}= - K~ \nabla T \bigg |_\vartheta,
\label{eq:1116}
\end{equation}

\begin{equation}
C~\frac {\partial T}{\partial \vartheta} + \nabla \cdot {\bf q}\bigg |_\vartheta = S (x,y,z,\vartheta),
\label{eq:1118}
\end{equation}

\begin{equation}
\frac {1}{\kappa}~\frac {\partial T}{\partial \vartheta} = \nabla^2 T \bigg |_\vartheta + \frac {1}{K}~ S (x,y,z,\vartheta),
\label{eq:1120}
\end{equation}

\noindent where ${\bf q},~T,~K,~C$ denote heat flux, temperature, heat conductivity, and specific heat capacity per unit volume of the continuum, respectively; $S(x,y,z,\vartheta)$ is a heat source term; $\kappa= K/C$ is thermal diffusivity. It is assumed that $K=const$. For infinite speed of the perturbations propagation, when $\vartheta=\tau$, equations (\ref{eq:1116})-(\ref{eq:1120}) become classical.

For regular independent  variables $\tau,~x,~y,~z$ and finite speed of the perturbations propagation, equations (\ref{eq:1116})-(\ref{eq:1120}) are transformed into 

\begin{equation}
{\bf q}= - K~ \bigg(\nabla T + \frac {\partial T}{\partial \tau}\cdot \nabla \tau_P \bigg)
\label{eq:1116aa}
\end{equation}

\begin{equation}
C~\frac {\partial T}{\partial \tau} + \nabla \cdot {\bf q}+ \frac {\partial {\bf q}}{\partial \tau} \cdot \nabla \tau_P = S (x,y,z,\vartheta),
\label{eq:1118aa}
\end{equation}

\begin{equation}
\frac {1}{\kappa}~\frac {\partial T}{\partial \tau} = \nabla^2 T + 2 \frac {\partial (\nabla T)}{\partial \tau}\cdot \nabla \tau_P+ \frac {\partial T}{\partial \tau} \nabla^2 \tau_P +\frac {\partial^2 T}{\partial \tau^2}(\nabla \tau_P)^2+\frac {1}{K}~ S (x,y,z,\tau)
\label{eq:1120aa}
\end{equation}

Equations (\ref{eq:1116aa}), (\ref{eq:1118aa}) and (\ref{eq:1120aa}) are identical to (\ref{eq:1116}), (\ref{eq:1118}) and (\ref{eq:1120}), respectively, and are quite different from classical Fourier equations predicting infinite speed of propagation. Only when $c_X \rightarrow \infty$, and therefore $\tau_P \rightarrow 0 $, they become classical. We see that using regular variables $\tau,~x,~y,~z$ instead of $\vartheta,~x,~y,~z$, makes equations describing transport process  with finite speed of the perturbations propagations, much more complicated and more difficult to solve.

In one dimensional case and for zero heat source term modified Fourier governing equation (\ref{eq:1120}) becomes 
\begin{equation}
\frac {\partial T}{\partial \vartheta} = \kappa \frac {\partial^2 T}{\partial x^2} \bigg |_\vartheta 
\label{eq:1120ab}
\end{equation}

\begin{figure}
\center
\includegraphics[width=12cm, height=10cm]{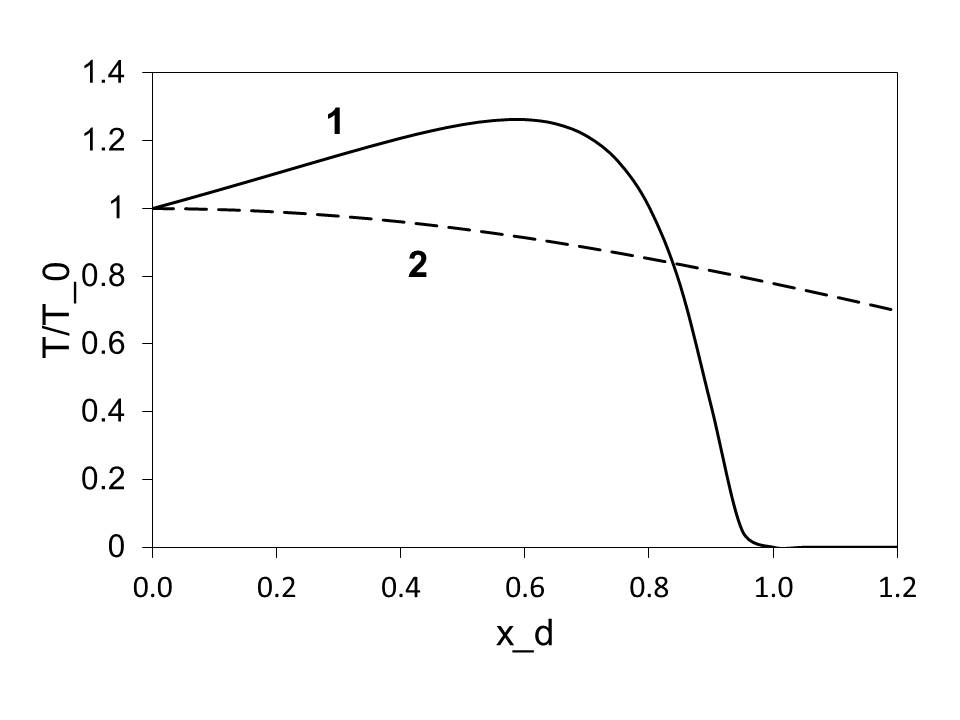}
\caption{Dimensionless temperature distribution from instantaneous plane heat source (1 - according to modified Fourier equation, 2 - according to classical Fourier equation).}
\end{figure}

We assume initial condition $T=0$ for $\vartheta=0$ and instantaneous source $Q$ at plane $x=0$ and $\vartheta=0$, while $Q~C$ is quantity of heat liberated per unit area of the plane. For this case solution of equation (\ref{eq:1120ab}) is
\begin{equation}
T=\frac {Q}{2 \sqrt{\pi \kappa \vartheta}}~\exp \bigg (-\frac {x^2}{4 \kappa \vartheta}\bigg )
\label{eq:1120ac}
\end{equation}
Expression (\ref{eq:1120ac}) is obtained by applying the local time concept to a respective solution of regular Fourier equation presented in \cite{sh8}.

As $\vartheta=\tau-\tau_P=\tau-|x|/c_X,~\tau=\tau_d \kappa/c_X^2,~|x|= x_d \kappa/c_X$, where $\tau_d,~x_d$ are dimensionless time and dimensionless absolute value of space coordinate, respectively, formula (\ref{eq:1120ac}) can be transformed into
\begin{equation}
\frac {T}{T_0}=\sqrt {\frac {\tau_d}{\tau_d-x_d}}~\exp \bigg[-\frac{x_d^2}{4 (\tau_d-x_d)} \bigg ]
\label{eq:1120ad}
\end{equation}
where $T_0$ is temperature at $x=0$.

Curve 1 in Fig. 1 is temperature distribution according to expression (\ref{eq:1120ad}) for $\tau_d=1$, and curve 2 is temperature distribution according to classical Fourier equation for the same value of $\tau_d=1$. We see that modified Fourier equation really describes heat conduction with finite speed of the perturbations propagation.

\section{Theoretical determination of speed of the perturbations propagation in transport processes}

In this section, we theoretically determine speed of the perturbations propagation $c_X$ in ideal gas, solid and plasma. It is assumed that ideal gas, Fermi electron gas in solid, and electrons and ions in plasma are in very close to equilibrium state. As a result, equilibrium distribution functions can be applied for calculations and analysis. It can be proven  that in this case there is
$|{\partial (\varphi f)}/{\partial X} |=A_c(X, Y, Z)~ B(T,v_X,v_Y,v_Z)$ where $T$ is temperature. Therefore formula (\ref{eq:1110}) for calculating speed of the perturbations propagation is transformed into an expression not containing space derivatives
\begin{equation}
c_X=\frac {\int v_X^{(+)}~B~dv_X^{(+)}~dv_Y~dv_Z} {\int B~dv_X^{(+)}~dv_Y~dv_Z}
\label{eq:1108ia1}
\end{equation}
For a case when $B(T,~v)$, where $v=\sqrt{v_X^2+v_Y^2+v_Z^2}$, and macroscopic velocity of the system is zero $\bf{v}_h=0$, expression (\ref{eq:1108ia1}) becomes simpler
\begin{equation}
c_X=\frac {\int v^3~B(v)~dv} {2 \int v^2~B(v)~dv}
\label{eq:1108ib}
\end{equation}

\subsection{Speed of the perturbations propagation in a moving media.}

Expression (\ref{eq:1110}) defines speed of the perturbations propagation in a general case. For a moving media, particles velocity $\bf v$  is a sum
\begin{equation}
{\bf v}={\bf v}_r+{\bf v}_h
\label{eq:1117c}
\end{equation}
where ${\bf v}_h(v_{hX},v_{hY},v_{hZ})$ is macroscopic velocity, in a case of fluid it is hydrodynamic velocity, and ${\bf v}_r(v_{rX},v_{rY},v_{rZ})$ is particles velocity in non-moving media when ${\bf v}_h=0$.

From expressions (\ref{eq:1110}) and (\ref{eq:1117c}) follows that
\begin{equation}
c_X=v_{hX}+\frac {I_a}{I_b}
\label{eq:1117d}
\end{equation}
\begin{equation}
I_a=\int^\infty_{v_{rX}=-v_{hX}} \int^\infty_{v_{rY}=-\infty} \int^\infty_{v_{rZ}= -\infty} v_{rX}~ \bigg |\frac{\partial (\varphi f)}{\partial X}\bigg |~dv_{rX}~dv_{rY}~dv_{rZ}
\label{eq:1117e}
\end{equation}
\begin{equation}
I_b=\int^\infty_{v_{rX}=-v_{hX}} \int^\infty_{v_{rY}=-\infty} \int^\infty_{v_{rZ}=-\infty} \bigg | \frac{\partial (\varphi f)}{\partial X}\bigg |~dv_{rX}~dv_{rY}~dv_{rZ}
\label{eq:1117f}
\end{equation}
where it is assumed that the distribution function $f$ depends only on ${\bf v}_r(v_{rX},v_{rY},v_{rZ})$ velocity components.

Expressions (\ref{eq:1117d})-(\ref{eq:1117f}) show that speed of the perturbations propagation increases when macroscopic velocity component $v_{hX}$ increases, and becomes zero only in a non-realistic case of $v_{hX} \rightarrow -\infty$. The function $c_X(v_{hX})$ has two asymptotes: at $v_{hX}>0$ a line $c_X=v_{hX}$ is an asymptote, and at $v_{hX}<0$ a line $c_X=0$ is the second asymptote. At small $v_{hX}$ there is
\begin{equation}
c_X \approx c_{0X}+ k_c v_{hX}
\label{eq:1117g}
\end{equation}
where $c_{0X}$ is speed of the perturbations propagations at $v_{hX}=0$, and $1>k_C>0$. Formula (\ref{eq:1117g}) shows that even for small values of macroscopic velocity, speed of the perturbations propagation in a moving media is not an additive function of the speed of the perturbations propagation in non-moving media and macroscopic velocity.

All these results regarding function $c_X(v_{hX})$ are obtained by using only two assumptions: (a) the improper integrals $I_a,~I_b$ converge, and (b) a function
\begin{equation}
I_c(v_{rX})=\int^\infty_{-\infty} \int^\infty_{-\infty} \bigg | \frac{\partial (\varphi f)}{\partial X}\bigg |~dv_{rY}~dv_{rZ}
\label{eq:1117h}
\end{equation}
does not depend on sign of particle velocity component $v_{rX}$.

\subsection{Isothermal viscous flow of ideal gas}

Speed of the perturbations propagation in isothermal ideal gas viscous flow is calculated using formulae (\ref{eq:1117d})-(\ref{eq:1117f}) and Maxwell distribution function from \cite{sh4}
\begin{equation}
f=n \bigg (\frac {m}{2 \pi k T}\bigg )^{\frac {3}{2}} \exp \bigg (- \frac {m v_r^2}{2 k T} \bigg )
\label{eq:1117i}
\end{equation}
where $n,~m,~k$ denote number of particles per unit volume, particle mass and the Boltzmann constant, respectively; $v_r^2=v_{rX}^2 + v_{rY}^2 + v_{rZ}^2$.

For viscous flow, the property of interest $\varphi$ is a vector of tangential momentum of a particle
\begin{equation}
\varphi=m~\bigg[(v_{hY}+v_{rY}) {\bf j} + (v_{hZ}+v_{rZ}) {\bf k} \bigg ]
\label{eq:1117k}
\end{equation}
where $\bf j,~\bf k$ are unit vectors parallel axes $Y$ and $Z$, respectively. As $v_{hY},~v_{hZ}$ and gas density $\rho=n~m$ are functions of space coordinates $X,Y,Z$ and time $\tau$, we get
\begin{equation}
\frac {\partial (\varphi f)}{\partial X}=\frac {\partial [\rho~(v_{hY} {\bf j} + v_{hZ} \bf k )]}{ \partial X} \bigg (\frac {m}{2 \pi k T}\bigg )^{\frac {3}{2}} \exp \bigg (- \frac {m v_r^2}{2 k T} \bigg )
\label{eq:1117l}
\end{equation}

\begin{figure}
\center
\includegraphics[width=12cm, height=10cm]{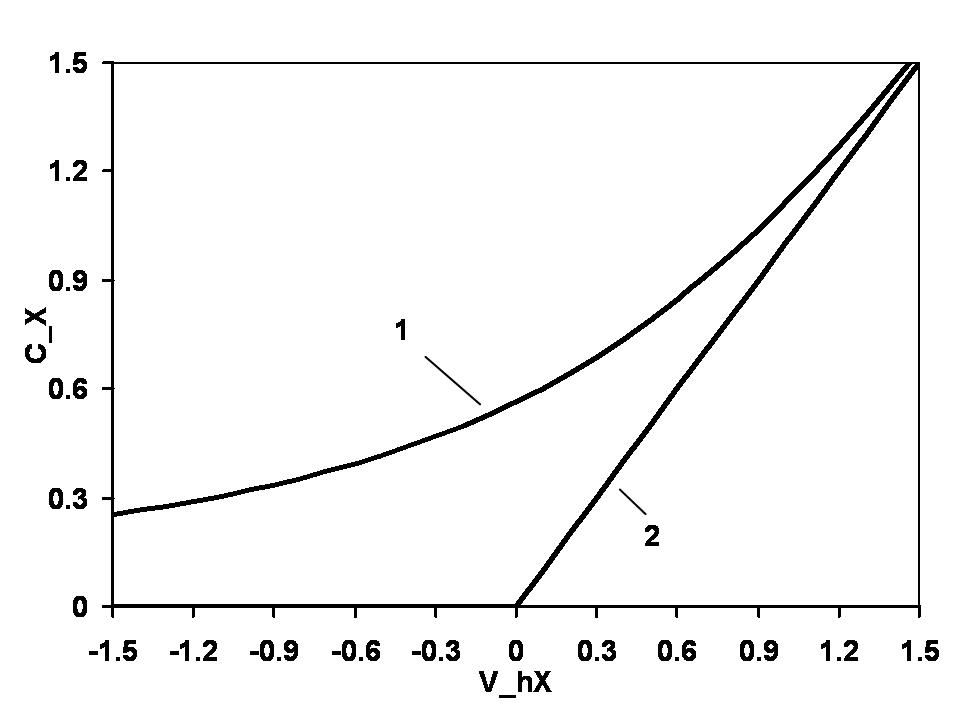}
\caption{Dimensionless speed of the perturbations propagation $C_X$ as a function of dimensionless hydrodynamic velocity normal component $V_{hX}$ for ideal gas isothermal viscous flow (1 - $C_X$, 2 - asymptote).}
\end{figure}

It is convenient to normalize all velocities using most probable particle speed $v_0=(2kT/m)^{0.5}$ from \cite{sh4} as a reference: $C_X=c_x/v_0,~V_{hX}=v_{hX}/v_0,~V_{rX}=v_{rX}/v_0,~V_{rY}=v_{rY}/v_0,
~V_{rZ}=v_{rZ}/v_0,~V_r=v_r/v_0$.

From expressions (\ref{eq:1108ia1}}), (\ref{eq:1117d})-(\ref{eq:1117f})  and (\ref{eq:1117l}) follows that dimensionless speed of the perturbations propagation in isothermal ideal gas viscous flow is 
\begin{equation}
C_X=V_{hX}+\frac {I_d}{I_e}=V_{hX}+\frac {1}{\sqrt {\pi}} \frac{\exp(-V_{hX}^2)}{ erfc(-V_{hX})}
\label{eq:1117m}
\end{equation}
where
\begin{displaymath}
I_d=\int^\infty_{V_{rX}=-V_{hX}} \int^\infty_{V_{rY}=-\infty} \int^\infty_{V_{rZ}= -\infty} V_{rX}~\exp (-V_r^2) ~dV_{rX}~dV_{rY}~dV_{rZ}=
\end{displaymath}
\begin{equation}
=\frac{\pi}{2}\exp(-V_{hX}^2)
\label{eq:1117n}
\end{equation}
\begin{displaymath}
I_e=\int^\infty_{V_{rX}=-V_{hX}} \int^\infty_{V_{rY}=-\infty} \int^\infty_{V_{rZ}=-\infty} \exp (-V_r^2) ~dV_{rX}~dV_{rY}~dV_{rZ}=
\end{displaymath}
\begin{equation}
=\frac {\pi \sqrt \pi}{2} erfc(-V_{hX})
\label{eq:1117o}
\end{equation}

Formula (\ref{eq:1117m}) and Fig. 2 confirm all previous general remarks regarding dependence of speed of the perturbations propagation from macroscopic velocity. For $V_{hX}=0$ speed of the perturbations propagation in isothermal viscous flow is simply
\begin{equation}
c_X=  \frac{1}{\sqrt \pi} \sqrt {\frac {2 kT}{m}}=0.56419 ~\sqrt {\frac {2 kT}{m}}
\label{eq:1117p}
\end{equation}

\subsection{Conduction of heat in ideal gas}

Speed of the perturbations propagation for a case of heat conduction in non-moving ideal gas is determined for an  arbitrary temperature distribution of  $T(X,Y,Z,\tau)$ and constant pressure conditions $p=const$. As $n=p/kT$, the distribution function (\ref{eq:1117i}) becomes
\begin{equation}
f=p \bigg (\frac {m}{2 \pi}\bigg )^{\frac {3}{2}} (kT)^{-5/2} \exp \bigg (- \frac {m v_r^2}{2 k T} \bigg )
\label{eq:1}
\end{equation}

The property of interest for conduction of heat is kinetic energy of a particle $\varphi=mv_r^2/2$. After simple calculations and by using introduced in the previous subsection dimensionless quantities we find that
\begin{displaymath}
\frac{\partial (\varphi f)}{\partial X}=A_c(X, Y, Z)~ B(V_r)
\end{displaymath}
where
\begin{displaymath}
A_c(X,Y,Z)=2.5 (m/2\pi k)^{1.5} T^{-2.5} \frac{\partial T}{\partial X},~ B(V_r)=V_r^2~(0.4~ V_r^2-1)~\exp(-V_r^2)
\end{displaymath}
It means that the integral mean value theorem does not hold, and formula (\ref{eq:1108ib}) is approximate. Numerical integration gives

\begin{equation}
C_X=\frac {I_f}{2~I_g}\approx 0.81755
\label{eq:2}
\end{equation}
\begin{equation}
I_f=\int_0^{\infty}V_r^5~|0.4~ V_r^2-1|~exp(-V_r^2)~dV_r \approx 5.30405
\label{eq:3}
\end{equation}
\begin{equation}
I_g=\int_0^{\infty}V_r^4~|0.4~ V_r^2-1|~exp(-V_r^2)~dV_r \approx 3.24387
\label{eq:4}
\end{equation}
and
\begin{equation}
c_X \approx 0.81755~ \sqrt {\frac {2 kT}{m}}
\label{eq:5}
\end{equation}

\subsection{Isothermal self-diffusion in ideal gas}

For determination of speed of the perturbations propagation for isothermal self-diffusion in non-moving ideal gas we assume an arbitrary space distribution of labeled particles quantity per unit volume $n(X,Y,Z)$ at $T=const$. As for labeled particles Maxwell distribution function holds, and property of interest is the particle velocity $\varphi=v_r$, we obtain
\begin{equation}
\frac{\partial (\varphi f)}{\partial X}=v_r \frac {\partial n}{\partial X} \bigg (\frac {m}{2 \pi k T}\bigg )^{\frac {3}{2}} \exp \bigg (- \frac {m v_r^2}{2 k T} \bigg )
\label{eq:6}
\end{equation}

Therefore the mean integral value theorem and relation (\ref{eq:1108ib}) are applicable, and dimensionless speed of the perturbations propagation is easily calculated
\begin{equation}
C_X=\frac {I_h}{2~I_i}=0.66467
\label{eq:7}
\end{equation}
where
\begin{equation}
I_h=\int_0^{\infty}V_r^4~exp(-V_r^2)~dV_r=\frac {1}{2}~\Gamma\bigg(\frac{5}{2}\bigg)
\label{eq:8}
\end{equation}
\begin{equation}
I_i=\int_0^{\infty}V_r^3~exp(-V_r^2)~dV_r=\frac {1}{2}~\Gamma(2)
\label{eq:9}
\end{equation}
Finally we get
\begin{equation}
c_X= 0.66467~ \sqrt {\frac {2 kT}{m}}
\label{eq:10}
\end{equation}
Integrals (\ref{eq:8}) and {\ref{eq:9}) are calculated using method described in \cite{sh39a}.

We see that speed of the perturbations propagation in ideal gas is different depending on type of transport process: at the same gas temperature, it is the highest for energy transport (thermal conductivity) at constant pressure, it is the lowest for momentum transport in isothermal viscous flow at $V_{hX}=0$, and it has intermediate value for mass transfer in isothermal self-diffusion.

\subsection{Conduction of heat in solid}

In dielectric crystals thermal perturbations propagate due to phonons, and because of this speed of the perturbations propagation is equal to speed of sound in the crystal if we take into account only acoustic phonons \cite{sh39a}. In metals thermal perturbations propagate in Fermi electron gas much faster then in lattice, because propagation speed in Fermi electron gas has an order of magnitude of Fermi velocity. Naturally, propagating thermal perturbations in electron gas cause induced local thermal perturbations in lattice. As a result, speed of the thermal perturbations propagation in metals is equal to speed of thermal perturbations propagation in Fermi electron gas, and the perturbations traveltime space distribution is the same in electron gas and lattice.

Fermi-Dirac distribution for electron gas at temperatures much lower than Fermi temperature is

\begin{equation}
f=\frac {2 m_e^3}{h^3} \frac {1} {\exp\bigg [r_F^{-1}(V^2-1)\bigg]+1}
\label{eq:11}
\end{equation}
where $V=v/v_F,~r_F=T/T_F,~v,~v_F,~m_e,~h,~T_F$ denote dimensionless velocity, dimensionless temperature, electron velocity, Fermi velocity, mass of electron, Plank constant and Fermi temperature, respectively.

Formula (\ref{eq:11}) is a modified version of classical expression presented in many sources, including for instance classical Born's book \cite{sh39b}, where is used Fermi-Dirac distribution function $f_e$ such that $dn_e=f_e~ \sqrt e ~de$ ($n_e$ and $e$ are number of electrons per unit volume and electron kinetic energy, respectively), while we use distribution function $f$ satisfying a condition  $dn_e=f~ dv_x~dv_y~dv_z$.

For heat conduction in electron gas, parameter of interest is $\varphi=m_e v^2/2$, so
\begin{displaymath}
\frac{\partial (\varphi f)}{\partial X}=A_c(X, Y, Z)~ B(V)
\end{displaymath}

\begin{figure}
\center
\includegraphics[width=12cm, height=10cm]{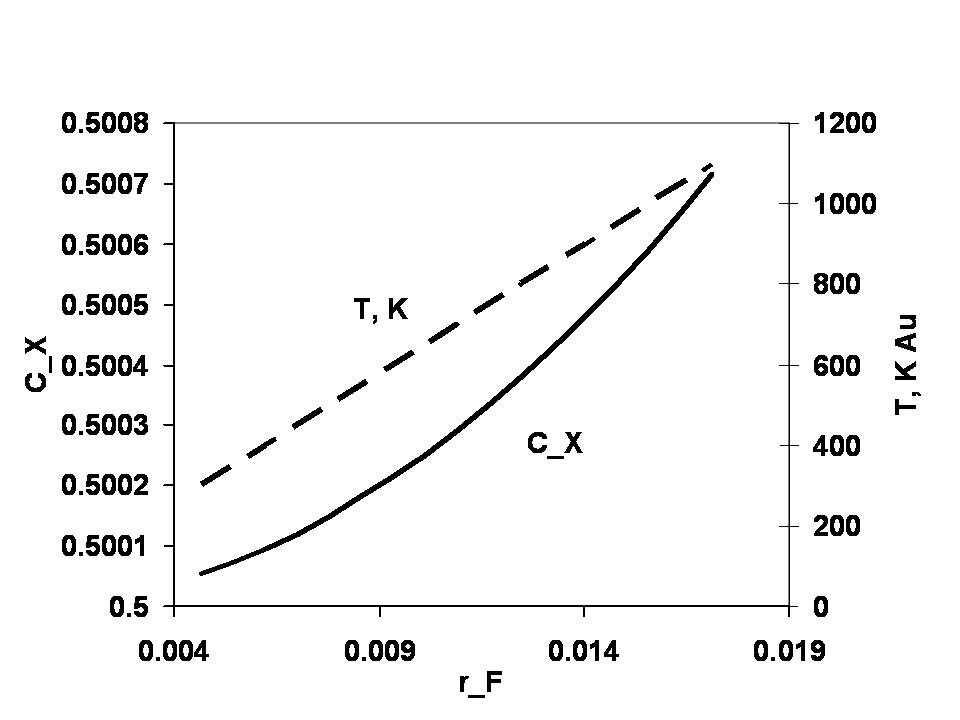}
\caption{Dimensionless speed of the perturbations propagation in electron gas $C_X$ and gold temperature as functions  of dimensionless temperature $r_F$.}
\end{figure}

where
\begin{displaymath}
A_c(X,Y,Z)= \frac {m_e^4 v_F^2 T_F}{h^3 T^2}  \frac{\partial T}{\partial X},~ B(V)=\frac {V^2 (V^2-1)\exp \bigg [r_F^{-1}(V^2-1)\bigg]} {\bigg[\exp \bigg (r_F^{-1}(V^2-1)\bigg)+1 \bigg]^2}
\end{displaymath}
It means that the integral mean value theorem is not applicable, and formula (\ref{eq:1108ib}) is approximate 

\begin{equation}
C_X=\frac{c_X}{v_F}=\frac {I_k}{2~I_l}
\label{eq:12}
\end{equation}
\begin{equation}
I_k=\int_0^\infty ~V^5 ~|V^2-1|~\exp \bigg (r_F^{-1}(V^2-1)\bigg)~ \bigg[\exp \bigg (r_F^{-1}(V^2-1)\bigg)+1 \bigg]^{-2}~dV
\label{eq:13}
\end{equation}
\begin{equation}
I_l=\int_0^\infty ~V^4 ~|V^2-1|~\exp \bigg (r_F^{-1}(V^2-1)\bigg)~ \bigg[\exp \bigg (r_F^{-1}(V^2-1)\bigg)+1 \bigg]^{-2}~dV
\label{eq:14}
\end{equation}

Fig. 3 presents values of $C_X$ obtained by numerical integration of expressions (\ref{eq:13}) and (\ref{eq:14}). It shows that in wide range of gold temperatures from 300 K to 1100 K speed of thermal perturbations propagation, practically, does not change, and its average value is $c_X=0.5003~ v_F$. As  $\lim_{r_F \rightarrow 0}~C_X=0.5$, we can conclude that speed of thermal perturbations propagation in Fermi gas and metals does not vary significantly in wide range of temperatures - from very low to moderate high.

\subsection{Conduction of heat in plasma}

Speed of the perturbations propagation in plasma is different depending on whether external magnetic field is present or not. If the external magnetic field is absent and plasma is close to equilibrium, Maxwell distribution can be applied for our calculations, and speed of the thermal perturbations propagation in plasma electron component is determined by approximate formula (\ref{eq:5}) where must be introduced mass of electron $m=m_e$. Mass of ions is much greater then mass of electrons $m_i \gg m_e$. However if we substitute $m=m_i$ in formula (\ref{eq:5}) we do not receive correct value of speed of the thermal perturbations propagation in plasma ion component because thermal perturbations, propagating in plasma electron component, cause induced  local thermal perturbations in the ion component. As a result, speed of the perturbations propagation in the ion component is equal to speed of the perturbations propagation in the electron component, and the perturbations traveltime space distributions are the same in both plasma components.

When external magnetic field $B_0$ exists in plasma, the most interesting is a case of the perturbations propagation in direction perpendicular to the field, because  conduction of heat in this direction strongly affects efficiency of nuclear fusion devices such as tokamaks and stellarators with magnetic confinement of plasma. In strong enough magnetic field,  gyrofrequencies of electrons and ions become much greater than their respective collisions frequencies.  In this case, from qualitative considerations presented in \cite{sh39c} follows that common for plasma electron and ion components speed of the thermal perturbations propagation  perpendicular to the field is
\begin{equation}
c_X \sim \frac{1}{B_0}
\label{eq:15a}
\end{equation}
So we can expect that in plasma with strong enough external magnetic field, speed of the thermal perturbations propagation perpendicular to the field is small. Experimental data presented in Ref. \cite{sh7a} supports this conclusion.

\section{Experimental verification of the perturbations propagation theory and the local time concept}

In this section, the perturbations propagation theory and the local time concept are tested using experimental data for conduction of heat in thin gold films and in hot fusion magnetically confined plasma.

\begin{figure}
\center
\includegraphics[width=12cm, height=10cm]{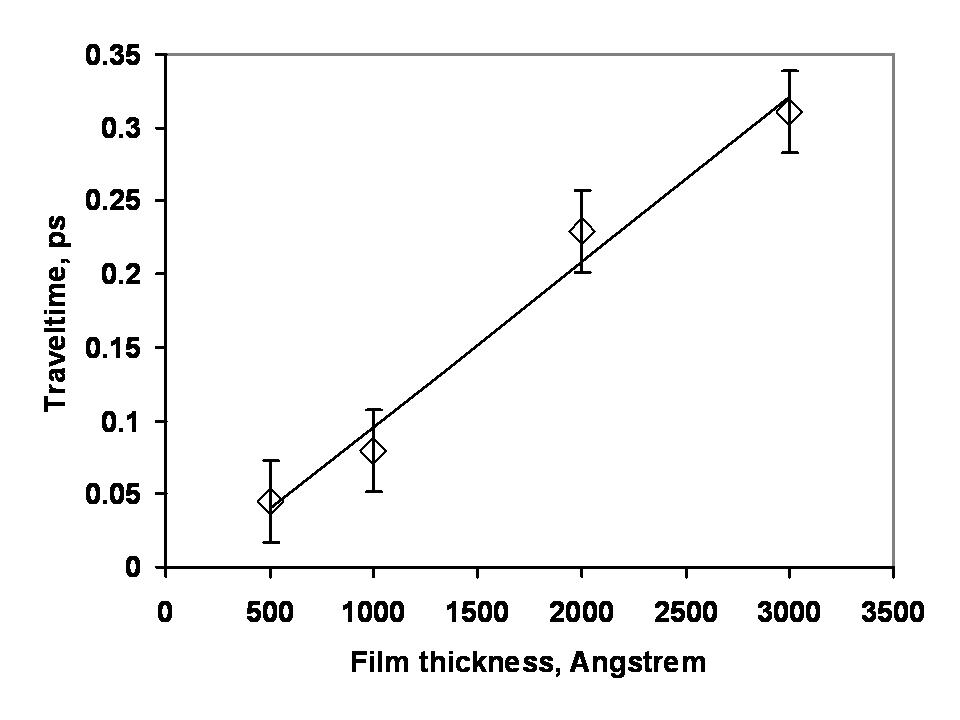}
\caption{Thermal pulses traveltime vs. gold films thickness according to works \cite{sh39d} and \cite{sh39e} presenting experimental results for femtosecond laser heating of thin gold films.}
\end{figure}

\subsection{Femtosecond laser heating of thin gold films}

Experimental data presented in works \cite{sh39d} and \cite{sh39e} makes possible to check validity of the local time concept for conduction of heat in metals, and also accuracy of our theoretical calculations relating to speed of the thermal perturbations propagation in metals. Authors of these works studied propagation of heat in thin gold films whose front surface was irradiated by very short femtosecond dye laser pulses with wavelength $\lambda_0 = 630\, nm$, pulse duration was $\tau_{lp} = 96\, fs$. They measured and registered the films back surface reflectivity history during and after irradiation. As the reflectivity change is caused by change of electron temperature, the cited experiments describe peculiarities of heat conduction at extremely short time intervals. Gold films thickness $L$ varied from 500 to 3,000 {\AA} and is significantly greater than gold optical skin depth $d_0 \approx 150$ {\AA}. In experiments, focal spot diameter of laser radiation on the film front surface is $d_s \approx 20,000$ {\AA} $\gg L$, therefore quasi one-dimensional temperature $T(x)$ and perturbation traveltime $\tau_p(x)$ distributions in a cylindrical control volume of diameter $d_c \ll d_s$ below the focal spot may be assumed, where $0 \le x \le L$ is a space coordinate along the control volume.

According to cited works \cite{sh39d} and \cite{sh39e}, change of films back surface reflectivity, which depends on electron temperature, starts with time delay $\tau_d$ after start of laser irradiation. It means that authors directly measured traveltime $\tau_p=\tau_d$ of the thermal perturbation. Their experimental data shows that it approximately linearly depends on the film thickness $L$ (Fig. 4). Using best fit (least square) approximation we calculated experimental dimensionless speed of the thermal perturbations propagation $C_X=c_x/v_F=(v_F~ d\tau_p/dL)^{-1}= 0.6 \pm 0.1$ and film thickness $L_0=146$ {\AA} corresponding to $\tau_p=0$. Thus, our theoretical value of $C_X = 0.5$ is in good agreement with experimental value of $C_X$. We see that $L_0 \approx d_0$, it means that in the region $0 \le x \le d_0$ there is $\tau_p=0$. This result is physically sound.

Experimental data presented in \cite{sh39d} and \cite{sh39e} makes possible to check the local time concept. For this purpose  we will use a system of coupled one-dimensional modified Fourier governing equations describing electron gas and lattice thermal interaction according to the local time concept
\begin{equation}
\frac {\partial T_e}{\partial \vartheta}=\kappa_e \frac {\partial^2 T_e}{\partial x^2}+g_e~(T_l-T_e) + a
\label{eq:15}
\end{equation}
\begin{equation}
\frac {\partial T_l}{\partial \vartheta}=\kappa_l \frac {\partial^2 T_l}{\partial x^2}+g_l~(T_e-T_l)
\label{eq:16}
\end{equation}
where indices $e,~l$ denote electron gas and lattice, respectively; $\kappa(T)=K(T)/C(T)$ is thermal diffusivity; $C(T),~K(T)$ are  heat capacity per unit volume and thermal conductivity, respectively; $g=G/C(T);~G$ is electron-phonon coupling constant; $a=A_l(\vartheta,x)/C_e(T_e),~A_l(\vartheta,x) $ is a heat source term resulting from laser beam absorption by electron gas.

For infinite speed of thermal perturbations propagation there is $\vartheta=\tau$, and equations (\ref{eq:15}) and (\ref{eq:16}) become regular two temperature model equations from \cite{sh39f} with $\kappa_l=0$, obviously not applicable in analysed case because finite speed of thermal perturbations propagation was observed in experiments.

The system of equations (\ref{eq:15})-(\ref{eq:16}) is nonlinear. Its numerical solution cannot give clear answer regarding applicability of the local time concept, because the works \cite{sh39d} and \cite{sh39e} do not present absolute values of the films back surface reflectivity which is affected by electron temperature. The only available experimental data, besides traveltime, is back surface reflectivity in arbitrary units vs. time counted from zero-delay point of laser beam. From these data we can determine only time moment $\tau_0(L)$ when extremum of reflectivity $\partial R/\partial \tau=0$ is observed. Since $R(T_e)$ is a monotonic function $\partial R/\partial T_e \neq 0$, and $\vartheta=\tau-\tau_p(L)$, we obtain $\partial R/\partial \tau=\partial R/\partial T_e~ \partial T_e/\partial \tau=\partial R/\partial T_e ~\partial T_e/\partial \vartheta=0$. Therefore extremum of reflectivity corresponds to maximum of temperature at local time moment $\vartheta_0(L)=\tau_0(L)-\tau_p(L)$. So we can check the local time concept, if we will find  analytical expression for $\vartheta_0(L)$ from the system of equations (\ref{eq:15})-(\ref{eq:16}).

\begin{figure}
\center
\includegraphics[width=12cm, height=10cm]{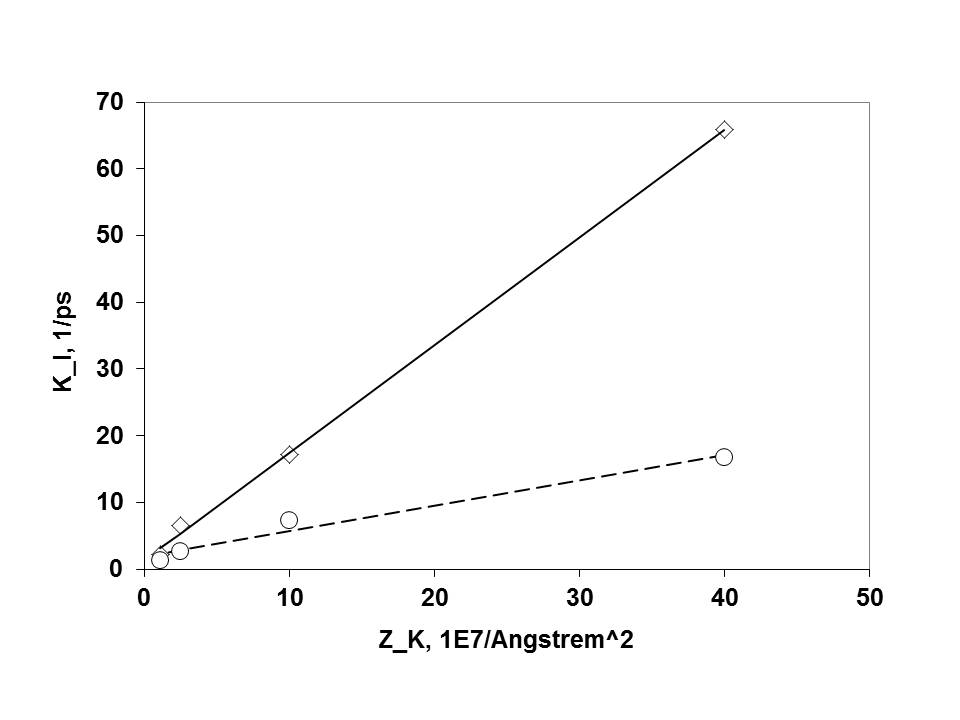}
\caption{Applying the local time concept (diamonds) and classical concept (circles) to experimental data from \cite{sh39d} and \cite{sh39e} for femtosecond laser heating of thin gold films.}
\end{figure}

In cited experiments the laser energy fluence was relatively small, approximately 1 mJ/cm$^2$. It results in moderate growth of the films temperature during laser heating and moderate changes of electron gas and lattice heat capacities. This allows to replace in equations (\ref{eq:15})-(\ref{eq:16}) instantaneous values of $C_e,~C_l,~\kappa_e,~\kappa_l$ by their average values without significant loss of accuracy. In this case, equations become linear with constant coefficients, and we can derive their analytical solutions $T_e(x, \vartheta),~T_l(x, \vartheta)$ corresponding to initial $T_e(x,0)=T_l(x,0)=T_0$ and boundary conditions $\partial T/\partial x(0,\vartheta)=\partial T/\partial x(L,\vartheta)=0$ which we can assume for analysed case of laser heating. After introducing excess temperatures $t_e=T_e-T_0,~t_l=T_l-T_0$ and applying separation of variables we get

\begin{equation}
t_e= \sum_{n=1}^{\infty} X_n(\vartheta)~cos\bigg(\frac{\pi n}{L} x \bigg)
\label{eq:17}
\end{equation}

\begin{equation}
t_l= \sum_{n=1}^{\infty} Y_n(\vartheta)~cos\bigg(\frac{\pi n}{L} x \bigg)
\label{eq:18}
\end{equation}

\begin{displaymath}
X_n(\vartheta)=\frac {1}{r_{1n}-r_{2n}} \int_0^{\vartheta}\bigg[(r_{1n}+m_n)~
\exp{(r_{1n}(\vartheta-\xi))} -
\end{displaymath}
\begin{equation}
-(r_{2n}+m_n)~\exp{(r_{2n}(\vartheta-\xi))} \bigg]a_n(\xi)~d\xi
\label{eq:19}
\end{equation}

\begin{equation}
Y_n(\vartheta)=\frac {g_l}{r_{1n}-r_{2n}} \int_0^{\vartheta}\bigg[\exp{(r_{1n}(\vartheta-\xi))} -
\exp{(r_{2n}(\vartheta-\xi))} \bigg]a_n(\xi)~d\xi
\label{eq:20}
\end{equation}
\begin{equation}
a_n(\vartheta)=\frac{2}{L}
\int_0^L a(\vartheta,\xi)~cos\bigg(\frac{\pi n}{L} \xi \bigg)~ d\xi
\label{eq:21}
\end{equation}
\begin{equation}
r_{1n,2n}=-\frac{m_n+k_n}{2} \pm \sqrt{\frac{(m_n-k_n)^2}{4}+g_e g_l}
\label{eq:22}
\end{equation}
\begin{equation}
k_n=\bigg( \frac{\pi n}{L} \bigg)^2 \kappa_e+g_e
\label{eq:23}
\end{equation}
\begin{equation}
m_n=\bigg( \frac{\pi n}{L}\bigg)^2 \kappa_l+g_l
\label{eq:24}
\end{equation}

In cited laser experiments with thin gold films there was $C_e \ll C_l,~ \kappa_e \gg \kappa_l,~g_e \gg g_l,~L/d_0>3$ and pulse duration $\tau_{lp}<1$ ps. For this conditions and an additional asymptotic condition $g_e \ll k_1$ the following asymptotic formula relating to electron temperature maximum is obtained from expressions (\ref{eq:17})-(\ref{eq:24})

\begin{equation}
K_l=\frac{A_K Z_K+g_e}{\theta_d}
\label{eq:25}
\end{equation}
\begin{equation}
K_l=\frac{1}{\vartheta_0(L)-\vartheta_{00}}
\label{eq:25a}
\end{equation}
\begin{equation}
Z_K= \frac{1}{L^2}
\label{eq:25b}
\end{equation}
\begin{equation}
A_K=\pi^2~\kappa_e
\label{eq:25c}
\end{equation}
where $\vartheta_{00}<\tau_{lp}$ and $\theta_d$ are parameters which depend on laser pulse duration and do not depend on the laser beam intensity.

Formulae (\ref{eq:25})-(\ref{eq:25c}) are corollaries of the local time concept. Expressions (\ref{eq:25})-(\ref{eq:25b}) include known from experiment values of $\vartheta_0(L)$ and $L$, and four unknown parameters $A_K,~d_e,~\vartheta_{00},~\theta_d$. They are determined by least squares $s=\sum [K_l-(A_K Z_K+d_e)/\theta_d]^2=\min$ after introducing experimental values of $\vartheta_0(L),~L$ in (\ref{eq:25})-(\ref{eq:25b}). Then can be calculated $\kappa_e=A_K/\pi^2,~T_e=G/g_e \gamma,~r_{\kappa}=\kappa_e/\kappa_{es}$ where $G=(3.0 \pm 0.5)\cdot 10^{16}~W/m^3 K,~\gamma=66~J/m^3 K^2 $ \cite{sh39g}, and thermal diffusivity of gold is $\kappa_{es}=1.57 \cdot 10^{-2}~m^2/s$. 

To make a comparison, the same calculational procedure can be applied for a classical case of infinite speed of the perturbations propagation if instead of experimental local time values $\vartheta_0(L)$ experimental global time values $\tau_0(L)$ are used.

Fig. 5 presents experimental and best fit values $K_l(Z_K)$ for local time   $\vartheta_0(L)$ and global time $\tau_0(L)$  coordinates of temperature maximums. Results of calculations using the local time concept are: $s=2.235~1/ps^2,~A_K=1.144 \cdot 10^7$ {\AA}$^2/ps,~g_e=0.917
~1/ps,~\theta_d=0.709,~\vartheta_{00}=0.0715~ps,
~\kappa_e=1.160 \cdot 10^{-2}~m^2/s,~T_e=496~K,~r_{\kappa}=0.74$. Results of calculations assuming infinite speed of the perturbations propagation are: $s=3.805~1/ps^2,~A_K=3.486 \cdot 10^6$ {\AA}$^2/ps,~ g_e=1.730~1/ps,~\theta_d=0.918,~\vartheta_{00}=0.0715~ps,
~\kappa_e=3.532\cdot 10^{-3}~m^2/s,~T_e=263 ~K,~r_{\kappa}=0.23$. For  a classical case was used the same value of $\vartheta_{00}=0.0715~ps$ as was determined by local time data because according to expressions (\ref{eq:25})-(\ref{eq:25b}) $\vartheta_{00}=\lim_{L \rightarrow 0} \vartheta_0(L)$. We could not calculate  experimental errors because authors of cited works did not provide enough data. 

We see that experimental data from \cite{sh39d} and \cite{sh39e} for femtosecond laser heating of thin gold films support the local time concept and reject classical approach. In the last case, variance $s$ is significantly greater, calculated from experimental data average electron temperature $T_e=263~K$ is too low taking into account that all experimental equipment had ambient initial temperature, and determined from experiment electron gas thermal diffusivity is too small. Application of the local time concept gave value of average electron temperature $T_e=496~K$ that is reasonably moderate and supports reasoning led to linearization of equations (\ref{eq:15})-(\ref{eq:16}), while obtained from femtosecond laser experiments value of electron gas thermal diffusivity $\kappa_e=1.160 \cdot 10^{-2}~m^2/s$ is close enough to $\kappa_{es}=1.57 \cdot 10^{-2}~m^2/s$.

So we can conclude that all available experimental data for femtosecond laser heating of thin gold films presented in \cite{sh39d} and \cite{sh39e} support formulated earlier theoretical predictions relating to speed of thermal perturbations propagation in metals and the local time concept.

\subsection{Power modulation experiments with JET tokamak}

In my  work  \cite{sh7a} the early versions of the perturbations propagation theory and the local time concept were verified for hot plasma by analyzis of experimental data for thermal pulses propagation in Alcator C-Mod tokamak. Now we will check validity of developed now theory and the local time concept for hot plasma in JET tokamak by analysing experimental data from paper \cite{sh39h}. Authors of this work applied modulated Ion Cyclotron Resonance Heating with 50/50 duty cycle, a modulation amplitude of approximately 80\% and different frequencies of modulation $f$. As the first step, we will use experimental data from \cite{sh39h} for plasma ion thermal diffusivities $\kappa_A(f) \sim (\partial \ln A_m / \partial R)^{-2},~\kappa_{\alpha}(f)\sim (\partial  \alpha / \partial R)^{-2}$ for $R \approx 3.4~m$ determined from modulated temperature amplitude $A_m$ and phase shift $\alpha$, respectively, where  $R$ is a linear coordinate in plasma along tokamak major radius. Authors of cited work found that $\kappa_A < \kappa_{\alpha}$. We will show that (a) the difference between $\kappa_A$ and  $\kappa_{\alpha}$ is a result of finite speed of the perturbations propagation in plasma, (b) it can be fully explained by the local time concept, (c) only $\kappa_A=\kappa$ is the real incremental thermal diffusivity.

In JET tokamak plasma vertical minor radius is 2.1 m, while plasma horizontal minor radius is 1.25 m only. Therefore along plasma major radius temperature field can be considered  approximately planar and one dimensional as it is usually assumed \cite{sh39i}. So we can apply periodical solution of modified according to the local time concept one-dimensional Fourier equation from our previous works \cite{sh7}, \cite{sh6}

\begin{equation}
A_m(R)=A_0~exp[-\xi~(R-R_0)]
\label{eq:26}
\end{equation}

\begin{equation}
\alpha(R)=-\bigg [\xi~(R-R_0)+\omega \bigg (\tau_P-\frac{n}{f}\bigg )\bigg ]
\label{eq:27}
\end{equation}

\begin{equation}
\xi= \frac {\sqrt 3}{2} \sqrt {\frac {\omega}{\kappa}}
\label{eq:28}
\end{equation}
where $\kappa,~ A_m,~A_0,~R_0,~\omega=2 \pi f$ denote incremental thermal diffusivity, modulated temperature amplitude, reference temperature amplitude, reference $R$ coordinate and cyclic frequency, respectively; $n=0,1,2,3....$. Formulae (\ref{eq:26})-(\ref{eq:27}) prove that $\kappa_A=\kappa$, and only $\kappa_A=\kappa$ is real incremental thermal diffusivity which is present in modified and classical Fourier equations. Formula (\ref{eq:27}) predicts that in periodical oscillating temperature fields perturbation traveltime $\tau_P$ includes an arbitrary constant $n/f$, while speed of thermal perturbations $c_X=|d \tau_p/d R|^{-1}$ does not depend on it.

Differentiation of formula (\ref{eq:27}) brings

\begin{equation}
\frac {d \alpha} {d R}=-\bigg (\xi + \frac {\omega}{c_R}\bigg )=-\xi_{\alpha}
\label{eq:29}
\end{equation}

\noindent where $c_R=(d \tau_P/d R)^{-1}$ is characteristic perturbation speed, and $\xi_{\alpha}=\sqrt 3/2~\sqrt {\omega /\kappa_{\alpha}}$. When in plasma thermal perturbations propagate outward, there is $c_R=c_X>0$. In the case of inward propagating perturbations there is $c_R=-c_X<0$. Therefore only at infinite speed of the perturbations propagation $\kappa_A =\kappa= \kappa_{\alpha}$. In reality, when always speed of the perturbations propagation is finite there is $\kappa_A= \kappa \neq \kappa_{\alpha}$.

From formula (\ref{eq:29}) follows 

\begin{equation}
c_R=\frac {2}{\sqrt 3}~ \frac {\omega^{0.5}}{\kappa_{\alpha}^{-0.5}-\kappa_A^{-0.5}}
\label{eq:30}
\end{equation}

We see that if $\kappa_{\alpha}>\kappa_A$, as measured by authors of  cited work \cite{sh39h}, then characteristic  propagation speed is negative $c_R<0$ (Fig. 6), and the perturbations propagate inward.  Fig. 6, built according experimental data from cited paper, shows that speed of the perturbations propagation increases when frequency of temperature oscillations grows. The data also confirms theoretical prediction that in magnetically confined fusion plasma, speed of the perturbations propagation perpendicular to the field is small. Thus, we can conclude that analysed experimental data from Ref. \cite{sh39h} support the theory of the perturbations propagation and the local time concept.

\begin{figure}
\center
\includegraphics[width=12cm, height=10cm]{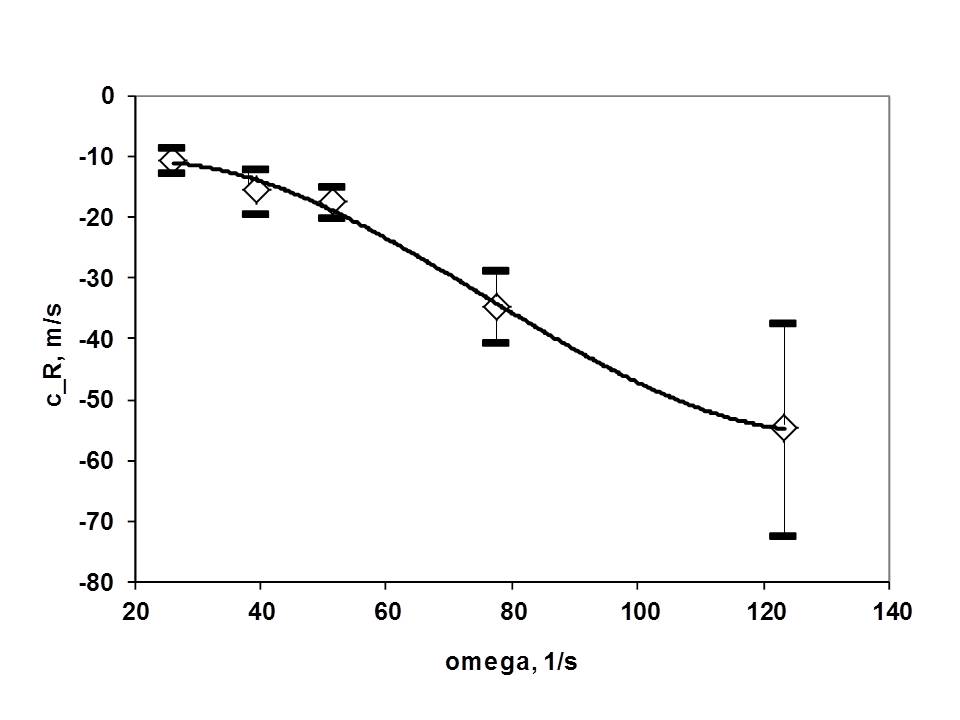}
\caption{Experimental values of characteristic perturbation speed $c_R$ of ion plasma component in JET tokamak vs. $\omega$ according to data from Ref. \cite{sh39h}.}
\end{figure}

The cited paper contains also results of simultaneous measurements of electron and ion components temperature amplitude and phase, though authors give experimental errors only for ion component. We use these experimental data for determination of simultaneous characteristic propagation speeds $c$ in ion and electron plasma components for $f=8~Hz,~3.2~m<R<3.6~m$. From experimental amplitude distributions $A(R)$ were calculated by least squares and formula (\ref{eq:26}) incremental thermal diffusivity, reference temperature amplitudes and reference $R$ coordinates  $\kappa_e=1.315~m^2/s,~\kappa_i=0.461~m^2/s,~A_{0e}=200.6~eV,~A_{0i}=165.6~eV,
~R_{0e}=3.07~m,~R_{0i}=3.18~m$ for electron and ion plasma components, respectively. Then formula (\ref{eq:27}) with fixed $n=0$ was applied to experimental phase distributions $\alpha(R)$ and were calculated experimental perturbation traveltime distributions $\tau_P(R)$ and, by numerical differentiation, experimental characteristic propagation speed distributions $c_R(R)=(d \tau_P/d R)^{-1}$ for ion and electron plasma components (Fig. 7). We see that (a) speed of thermal perturbations propagation in JET tokamak perpendicular to magnetic field is small and depends on $R$, most probably because of varying magnetic field $B_0(R)$ as it follows from formula (\ref{eq:15a}); (b) for electron and ion components experimental characteristic propagation speeds $c_R(R)$ have the same negative sign; (c) determined values of experimental characteristic propagation speed $c_R(R)$ for ion component are close to experimental characteristic propagation speed for electron component. Therefore we can conclude that results of this check support the local time concept and do not contradict our theoretical suggestion, that speed of the perturbations propagation and the thermal perturbations traveltime space distributions in the electron and ion plasma components are the same.

\begin{figure}
\center
\includegraphics[width=12cm, height=10cm]{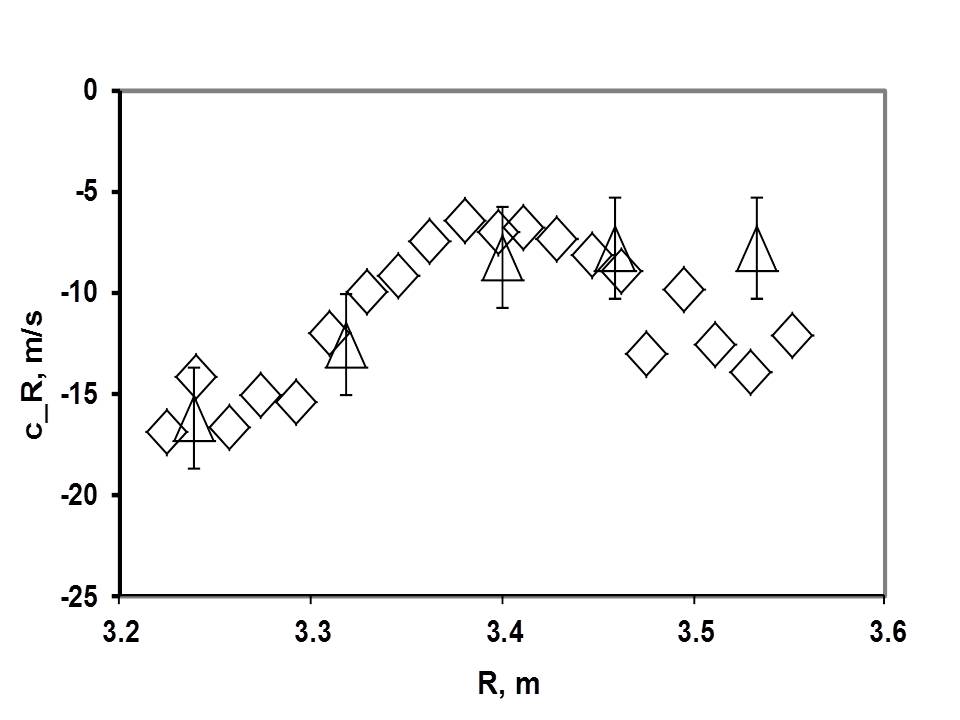}
\caption{Simultaneous space distributions of experimental characteristic perturbation speed $c_R$ in ion (triangles) and electron (diamonds) plasma components of JET tokamak according to data from Ref. \cite{sh39h}.}
\end{figure}

\section{Conclusions}

\begin{enumerate}

\item Developed general statistical theory of the perturbations propagation and the local time concept  are valid for  all non-equilibrium systems in gases, liquids, solids and plasma with any macroscopic velocities.

\item Finite speed of the perturbations propagation is predicted for all  transport processes. It depends on the system properties, the process type and macroscopic velocity component normal to the boundary separating perturbed and non-perturbed subregions.

\item Propagation of perturbations in transport processes is described by primary waves solutions of derived eikonal type equation for the perturbations traveltime.

\item Developed in the work local time concept allows to formulate  modified kinetic, conservation and governing equations describing real transport processes with finite speed of the perturbations propagation.

\item At the same temperature, speed of the perturbations propagation in ideal gas is the highest for energy transport (thermal conductivity) at constant pressure, it is the lowest for momentum transport in isothermal viscous flow with zero normal macroscopic velocity component, and it has intermediate value for mass transfer in isothermal self-diffusion.

\item For metals, speed of thermal perturbations propagation is the same in Fermi electron gas and lattice. At low, ambient and  moderate high temperature, it is, approximately, half of Fermi velocity. In dielectric crystals, speed of thermal perturbations propagation is equal to speed of sound in the crystal.

\item In plasma, speed of thermal perturbations  propagation is the same in electron and ion components. In a case of strong enough external magnetic field, speed of thermal perturbations propagation perpendicular to the field is small.

\item Published experimental data for femtosecond laser heating of thin gold films is in satisfactory agreement with modified coupled Fourier governing equations, resulting from the local time concept, and calculated theoretical speed of thermal perturbations propagation in metals.

\item Published results of power modulation experiments in JET tokamak are in satisfactory agreement with modified Fourier governing equation, obtained from the local time concept, theoretical prediction of small thermal perturbations propagation speed perpendicular external strong magnetic field, and support our theoretical suggestion  that speed of the perturbations propagation and the thermal perturbations traveltime space distributions in the electron and ion plasma components are the same.

\end{enumerate}

\end{document}